\documentclass[conference]{IEEEtran}
\IEEEoverridecommandlockouts
\usepackage{cite}
\usepackage{amsmath,amssymb,amsfonts}
\usepackage{algorithmic}
\usepackage{graphicx}
\usepackage{textcomp}
\usepackage{xcolor}
\def\BibTeX{{\rm B\kern-.05em{\sc i\kern-.025em b}\kern-.08em
    T\kern-.1667em\lower.7ex\hbox{E}\kern-.125emX}}

\usepackage[caption=false, font=footnotesize]{subfig}

\graphicspath{{fig/}}

\begin{document}

\title{Channel Measurement for Holographic MIMO: Benefits and Challenges of Spatial Oversampling
}

\author{Tengjiao Wang\IEEEauthorrefmark{1}, Yongxi Liu\IEEEauthorrefmark{2}, Ming Zhang\IEEEauthorrefmark{2}, Wei E. I. Sha\IEEEauthorrefmark{3}, Cen Ling\IEEEauthorrefmark{1}, Chao Li\IEEEauthorrefmark{1},  Shaobo Wang\IEEEauthorrefmark{1}

\\ \IEEEauthorrefmark{1}Wireless Network RAN Research Department, Huawei Technologies CO., Ltd, Shanghai, China 
\\ \IEEEauthorrefmark{2}School of Electronic and Information Engineering, Xi'an Jiaotong University, Shaanxi, China
\\ \IEEEauthorrefmark{3}College of Information Science and Electronic Engineering, Zhejiang University, Hangzhou, China
\\ Emails: \IEEEauthorrefmark{1}\{wangtengjiao6, lingcen, lichao18, shaobo.wang\}@huawei.com,
\\ \IEEEauthorrefmark{2}liuyongxi@stu.xjtu.edu.cn, ming20.zhang@xjtu.edu.cn, \IEEEauthorrefmark{3} weisha@zju.edu.cn
}

\maketitle

\begin{abstract}
In this paper, the channel of an indoor holographic multiple-input multiple-output (MIMO) system is measured. It is demonstrated through experiments for the first time that the spatial oversampling of holographic MIMO systems is able to increase the capacity of a wireless communication system significantly. However, the antenna efficiency is the most crucial challenge preventing us from getting the capacity improvement. An extended EM-compliant channel model is also proposed for holographic MIMO systems, which is able to take the non-isotropic characteristics of the propagation environment, the antenna pattern distortion, the antenna efficiency, and the polarization characteristics into consideration.
\end{abstract}


\section{Introduction} 
In order to increase the spectral efficiency and energy efficiency for the future 5G-Advanced and 6G wireless communication systems, the concept of holographic multiple-input multiple-output (MIMO) is proposed recently~\cite{2020-Chongwen-Holo}. By integrating an infinite number of antennas into a limited surface, holographic MIMO is expected to fully exploit the propagation characteristics offered by the electromagnetic channel and approach the fundamental performance limit~\cite{2021-Dardari-Holo}. For holographic MIMO systems, both the accurate channel modeling and the realistic performance evaluation are key problems.

In the literature, many efforts have been devoted to accurately model the channel of holographic MIMO systems~\cite{2020-Pizzo-HoloChannel, 2022-Chongwen-HoloChannel, 2022-Pizzo-HoloChannel,2022-Tengjiao-Holo, 2022-Yongxi-Holo}. In~\cite{2020-Pizzo-HoloChannel}, a Fourier plane-wave series expansion-based channel model is proposed for holographic MIMO systems. Based on the Fourier spectral representation, it provides a physically meaningful model capturing the propagation characteristics of the electromagnetic (EM) wave. In~\cite{2022-Chongwen-HoloChannel}, the authors extend the Fourier plane-wave channel model to a multi-user scenario. In~\cite{2022-Pizzo-HoloChannel}, the non-isotropic scattering environment is further considered by using the von Mises-Fisher (VMF) distributions~\cite{2009-Mammasis-Mises}. Then in our previous works~\cite{2022-Tengjiao-Holo,2022-Yongxi-Holo}, an EM-compliant channel model is proposed. By combining the VMF distributions~\cite{2009-Mammasis-Mises} and the 3GPP TR 38.901 channel model~\cite{2020-Standard-38901}, a realistic angular power spectrum is modeled. The non-ideal factors caused by mutual coupling at the transceivers~\cite{2015-Balanis-Antenna}, including the antenna pattern distortion and the decrease of antenna efficiency are also accounted for. However, in the state-of-the-art channel models~\cite{2020-Pizzo-HoloChannel, 2022-Chongwen-HoloChannel, 2022-Pizzo-HoloChannel,2022-Tengjiao-Holo, 2022-Yongxi-Holo}, the polarization of EM wave is not taken into consideration. The polarization is an inherent characteristic of EM waves and will have significant impact on the performance of holographic MIMO systems.

Another key problem for holographic MIMO is the accurate performance evaluation. In~\cite{2022-Pizzo-HoloChannel}, the ergodic capacity of a single-user holographic MIMO system is analyzed, which shows the capacity improvement of holographic MIMO over conventional MIMO. In~\cite{2021-Linglong-CAP}, the capacity of a single-user holographic MIMO system is investigated from a continuous point of view, the capacity enhancement is also demonstrated. In our previous work~\cite{2022-Tengjiao-Holo}, both the single-user and the multi-user downlink channel capacities of holographic MIMO systems are investigated. However, the performance evaluations in the state-of-the-art researches~\cite{2022-Pizzo-HoloChannel, 2022-Tengjiao-Holo, 2021-Linglong-CAP} are all  numerical simulations based on theoretical models. No experiments have been done to verify the accuracy of the performance evaluations.

In this paper, we try to solve the above two problems for holographic MIMO systems. Firstly, an extended EM-compliant channel model is proposed based on the channel model in our previous work~\cite{2022-Tengjiao-Holo}. Secondly, the channel of holographic MIMO is measured through real-world experiments, and the performance is evaluated based on the measurement results. The contributions of this paper can be summarized as follows:
\begin{itemize}
\item An extended EM-compliant channel model is proposed for holographic MIMO systems. In the extended model, not only the non-isotropic characteristics of the propagation environment, the antenna pattern distortion, the antenna efficiency, but also the polarization of the antennas and the propagation environment can be modeled. 

\item Based on the extended channel model, the real-world channel for holographic MIMO systems is measured for the first time in an indoor environment. An experiment is carefully designed, in which an electrically controlled virtual dense array is used to realize arbitrary element spacings of holographic MIMO.	

\item The channel capacity of holographic MIMO systems is evaluated according to the measurement results. It is demonstrated that the spatial oversampling of holographic MIMO is able to provide a two to three times capacity enhancement, without considering the antenna efficiency loss. The antenna efficiency is the most crucial challenge for holographic MIMO.

\end{itemize}

The rest of this paper is organized as follows. In Section~\ref{Sec-Holo}, the proposed extended EM-compliant channel model for holographic MIMO is explained in details. Then, the setup for the channel measurement is given in Section~\ref{Sec-Setu}. The measurement results and the corresponding performance evaluations are provided in Section~\ref{Sec-Simu}. Finally, this paper is concluded in Section~\ref{Sec-Conc}.

\section{Extended EM-Compliant Channel Model}  \label{Sec-Holo}
\subsection{EM-Compliant Channel Model}
In this subsection, the EM-compliant channel model for holographic MIMO in our previous work~\cite{2022-Tengjiao-Holo} is briefly introduced. The central frequency and wavelength are denoted by $f_c$ and $\lambda$. Planar antenna arrays with size $\{ L_\mathrm{R}^x, L_\mathrm{R}^y \}$ and $\{ L_\mathrm{S}^x, L_\mathrm{S}^y \}$ are equipped at the receiver and the transmitter, respectively. The numbers of antenna elements are denoted by $N_\mathrm{R}$ and $N_\mathrm{S}$. The spacings between antenna elements are denoted by $\{ \Delta_\mathrm{R}^x, \Delta_\mathrm{R}^y \}$ and $\{ \Delta_\mathrm{S}^x, \Delta_\mathrm{S}^y \}$. The coordinates of the antenna elements are represented by $\mathbf{r}_q = (r_q^x, r_q^y, r_q^z), q = 1,2,\cdots,N_\mathrm{R}$ and $\mathbf{s}_p = (s_p^x, s_p^y, s_p^z), p = 1,2,\cdots,N_\mathrm{S}$, respectively. 

According to~\cite{2022-Tengjiao-Holo}, the channel matrix $\mathbf{H} \in \mathbb{C}^{N_\mathrm{R} \times N_\mathrm{S}}$ can be expressed as
\begin{equation} \label{H_final}
	\mathbf{H} = \boldsymbol{\Gamma}_\mathrm{R} \boldsymbol{\Psi}_\mathrm{R} {\mathbf{H}}_\mathrm{a} \boldsymbol{\Psi}_\mathrm{S}^\mathrm{H} \boldsymbol{\Gamma}_\mathrm{S},
\end{equation}
where ${\mathbf{H}}_\mathrm{a} \in \mathbb{C}^{n_\mathrm{R} \times n_\mathrm{S}}$ denotes the wavenumber-domain channel matrix, which has $n_\mathrm{R} \times n_\mathrm{S}$ elements. Here, $n_\mathrm{R} =  | \mathcal{E}_\mathrm{R} |$ and $n_\mathrm{S} = |\mathcal{E}_\mathrm{S}|$ are the cardinalities of the sets $\mathcal{E}_\mathrm{R}$ and $\mathcal{E}_\mathrm{S}$, with $\mathcal{E}_\mathrm{R} = \left\{ (l^x, l^y) \in \mathbb{Z}^2: \left( \frac{l^x \lambda}{L_\mathrm{R}^x} \right)^2 + \left( \frac{l^y \lambda}{L_\mathrm{R}^y} \right)^2 \leq 1 \right\}$ and $\mathcal{E}_\mathrm{S} = \left\{ (m^x, m^y) \in \mathbb{Z}^2: \left( \frac{m^x \lambda}{L_\mathrm{S}^x} \right)^2 + \left( \frac{m^y \lambda}{L_\mathrm{S}^y} \right)^2 \leq 1 \right\}$. Each element $[\mathbf{H}_\mathrm{a}]_{\beta, \alpha}$ of ${\mathbf{H}}_\mathrm{a}$ is a random Fourier coefficient following the complex Gaussian distribution $\mathcal{CN}(0, \sigma^2_{\beta, \alpha}), \beta = 1,2,\cdots,n_\mathrm{R}, \alpha = 1,2,\cdots,n_\mathrm{S}$. The variance can be further given by
\begin{equation}
\begin{split}
	\sigma^2_{\beta, \alpha} = & \int \!\!\! \int_{\Omega_\mathrm{R}(l^x_{\beta}, l^y_{\beta})} \!\!\! A^2(\theta_\mathrm{R}, \phi_\mathrm{R}) \sin \theta_\mathrm{R} \mathrm{d}\theta_\mathrm{R} \mathrm{d}\phi_\mathrm{R} \times \\
& \int \!\!\! \int_{\Omega_\mathrm{S}(m^x_\alpha, m^y_\alpha)} \!\!\! A^2(\theta_\mathrm{S}, \phi_\mathrm{S}) \sin \theta_\mathrm{S} \mathrm{d}\theta_\mathrm{S} \mathrm{d}\phi_\mathrm{S},
\end{split}
\end{equation}
where $A^2(\theta_\mathrm{R}, \phi_\mathrm{R})$ and $A^2(\theta_\mathrm{S}, \phi_\mathrm{S})$ denote the angular power spectrum at the receiver and the transmitter, respectively. The angular power spectrum can be further modeled by a mixture of VMF distributions~\cite{2009-Mammasis-Mises}
\begin{equation}
	A^2_\mathrm{R} (\theta_\mathrm{R}, \phi_\mathrm{R}) = \sum_{i=1}^{N_c} w_{\mathrm{R},i} p_{\mathrm{R},i}(\theta_\mathrm{R}, \phi_\mathrm{R}),
\end{equation}
and
\begin{equation}
	A^2_\mathrm{S} (\theta_\mathrm{S}, \phi_\mathrm{S}) = \sum_{i=1}^{N_c} w_{\mathrm{S},i} p_{\mathrm{S},i}(\theta_\mathrm{S}, \phi_\mathrm{S}),
\end{equation}
where $N_c$ denotes the number of clusters of the scatters in the propagation environment. $w_{\mathrm{R},i}$ and $w_{\mathrm{S},i}$ denote the normalization factor with $\sum_i^{N_c} w_{\mathrm{R},i} = \sum_i^{N_c} w_{\mathrm{S},i} =  1$. $p_{\mathrm{R},i}(\theta_\mathrm{R}, \phi_\mathrm{R})$ and $p_{\mathrm{S},i}(\theta_\mathrm{S}, \phi_\mathrm{S})$ denote the probability functions of the VMF distribution, which can be further expressed as~\cite{2009-Mammasis-Mises}
\begin{equation}
\begin{split}
	p_{\mathrm{R},i}&(\theta_\mathrm{R}, \phi_\mathrm{R}) = \frac{\alpha_{\mathrm{R},i}}{4 \pi \mathrm{sinh} (\alpha_{\mathrm{R},i})} \times \\
& e^{\alpha_{\mathrm{R},i}(\sin \theta_\mathrm{R} \sin \bar{\theta}_{\mathrm{R},i} \cos(\phi_\mathrm{R} - \bar{\phi}_{\mathrm{R},i}) + \cos \theta_\mathrm{R} \cos \bar{\theta}_{\mathrm{R},i})},
\end{split}
\end{equation}
and
\begin{equation}
	\begin{split}
		p_{\mathrm{S},i}&(\theta_\mathrm{S}, \phi_\mathrm{S}) = \frac{\alpha_{\mathrm{S},i}}{4 \pi \mathrm{sinh} (\alpha_{\mathrm{S},i})} \times \\
	& e^{\alpha_{\mathrm{S},i}(\sin \theta_\mathrm{S} \sin \bar{\theta}_{\mathrm{S},i} \cos(\phi_\mathrm{S} - \bar{\phi}_{\mathrm{S},i}) + \cos \theta_\mathrm{S} \cos \bar{\theta}_{\mathrm{S},i})},
	\end{split}
\end{equation}
where $\{\bar{\phi}_{\mathrm{R},i}, \bar{\theta}_{\mathrm{R},i}\}$ and $\{\bar{\phi}_{\mathrm{S},i}, \bar{\theta}_{\mathrm{S},i}\}$ denote the elevation and azimuth angles of the $i$-th cluster at the receiver and the transmitter. $\alpha_{\mathrm{S},i}$ and $\alpha_{\mathrm{R},i}$ denote the concentration parameters for the $i$-th cluster. These angles can be derived from the 3GPP TR 38.901 channel model~\cite{2020-Standard-38901} according to the relationship defined in~\cite{2022-Tengjiao-Holo}.

In~\eqref{H_final}, $\boldsymbol{\Psi}_\mathrm{R} \in \mathbb{C}^{N_\mathrm{R} \times n_\mathrm{R}}$ and $\boldsymbol{\Psi}_\mathrm{S} \in \mathbb{C}^{N_\mathrm{S} \times n_\mathrm{S}}$ denote the modified Fourier harmonics, which take the antenna pattern distortion into consideration. Each element of $\boldsymbol{\Psi}_\mathrm{R}$ and $\boldsymbol{\Psi}_\mathrm{S}$ can be further derived as

\begin{equation} \label{equ_phi_R}
\begin{split}
	[\boldsymbol{\Psi}_\mathrm{R}]_{q, \beta} =
	& \frac{1}{\sqrt{N_\mathrm{R}}} e^{j\left( \frac{2\pi l^x_\beta}{L_\mathrm{R}^x} r_q^x + \frac{2\pi l^y_\beta}{L_\mathrm{R}^y} r_q^y + \gamma_\mathrm{R}(l^x_\beta,l^y_\beta) r_q^z\right)} \\
	& \times F_{\mathrm{R},q}\left( \hat{\theta}_\mathrm{R}(l^x_\beta, l^y_\beta) , \hat{\phi}_\mathrm{R}(l^x_\beta, l^y_\beta) \right),
\end{split}
\end{equation}
and
\begin{equation} \label{equ_phi_S}
\begin{split}
	[\boldsymbol{\Psi}_\mathrm{S}]_{p, \alpha} =
	 &\frac{1}{\sqrt{N_\mathrm{S}}} e^{j\left( \frac{2\pi m^x_\alpha}{L_\mathrm{S}^x} s_p^x + \frac{2\pi m^y_\alpha}{L_\mathrm{S}^y} s_p^y + \gamma_\mathrm{S}(m^x_\alpha,m^y_\alpha) s_p^z\right)} \\
 	& \times F_{\mathrm{S},p}\left( \hat{\theta}_\mathrm{S}(m^x_\alpha, m^y_\alpha) , \hat{\phi}_\mathrm{S}(m^x_\alpha, m^y_\alpha) \right),
\end{split}
\end{equation}
where $\gamma_\mathrm{R}(l^x, l^y) = \sqrt{(\frac{2\pi}{\lambda})^2 - (\frac{2\pi l^x}{L_\mathrm{R}^x})^2 - (\frac{2\pi l^y}{L_\mathrm{R}^y})^2}$ and $\gamma_\mathrm{S}(m^x, m^y) = \sqrt{(\frac{2\pi}{\lambda})^2 - (\frac{2\pi m^x}{L_\mathrm{S}^x})^2 - (\frac{2\pi m^y}{L_\mathrm{S}^y})^2}$. $F_{\mathrm{R},q}\left( \theta_\mathrm{R}, \phi_\mathrm{R} \right)$ and $F_{\mathrm{S},p}\left( \theta_\mathrm{S}, \phi_\mathrm{S} \right)$ represent the embedded element directivity pattern of the $q$-th antenna at the receiver and the $p$-th antenna at the transmitter. The corresponding elevation and azimuth angles $\{ \hat{\phi}_\mathrm{R}(l^x, l^y) , \hat{\theta}_\mathrm{R}(l^x, l^y) \}$ and $\{ \hat{\phi}_\mathrm{S}(m^x, m^y) , \hat{\theta}_\mathrm{S}(m^x, m^y) \}$ for the Fourier harmonics $(l^x, l^y)$ and $(m^x, m^y)$ can be calculated by a transformation from the wavenumber domain to the angular domain~\cite{2022-Tengjiao-Holo}.

In the end, $\boldsymbol{\Gamma}_\mathrm{R} \in \mathbb{R}^{N_\mathrm{R} \times N_\mathrm{R}}$ and $\boldsymbol{\Gamma}_\mathrm{S} \in \mathbb{R}^{N_\mathrm{S} \times N_\mathrm{S}}$ are diagonal matrices representing the efficiency of the antenna element at the receiver and the transmitter, respectively. More details of the channel model can be found in~\cite{2022-Tengjiao-Holo}. However, the polarization characteristics of the antenna and the environment are not taken into consideration.

\subsection{Extended Channel Model with Polarization}
In this subsection, we extend the EM-compliant channel model to account for the polarization characteristics of the antennas and the propagation environment.

From~\eqref{H_final}, the channel between the $p$-th antenna at the transmitter and the $q$-th antenna at the receiver can be expressed as
\begin{equation}
	[\mathbf{H}]_{q,p} = \sum_{\beta=1}^{n_\mathrm{R}} \sum_{\alpha=1}^{n_\mathrm{S}} \eta_{\mathrm{R},q} [\boldsymbol{\Psi}_\mathrm{R}]_{q,\beta} [\mathbf{H}_\mathrm{a}]_{\beta,\alpha} [\boldsymbol{\Psi}_\mathrm{S}]_{p,\alpha}^* \eta_{\mathrm{S},p},
\end{equation}
where $\eta_{\mathrm{R},q}$ and $\eta_{\mathrm{S},p}$ denote the diagonal elements of $\boldsymbol{\Gamma}_\mathrm{R}$ and $\boldsymbol{\Gamma}_\mathrm{S}$, representing the antenna efficiency of the corresponding antenna. According to~\cite{2015-Balanis-Antenna}, the polarization pattern of an EM wave can be decomposed into two components orthogonal to the propagation direction, i.e., the vertical polarization and the horizontal polarization. Therefore, the channel considering the polarization characteristics can be written as
\begin{equation} \label{equ-chn-pol-item}
	\begin{split}
		[\mathbf{H}]_{q,p}^{\mathrm{pol}} = 
		&\sum_{\beta=1}^{n_\mathrm{R}} \sum_{\alpha=1}^{n_\mathrm{S}} \eta_{\mathrm{R},q} [\boldsymbol{\Psi}_\mathrm{R}^\theta]_{q,\beta} [\mathbf{H}_\mathrm{a}^{\theta\theta}]_{\beta,\alpha} [\boldsymbol{\Psi}_\mathrm{S}^\theta]_{p,\alpha}^* \eta_{\mathrm{S},p} \\
		&+ \sum_{\beta=1}^{n_\mathrm{R}} \sum_{\alpha=1}^{n_\mathrm{S}} \eta_{\mathrm{R},q} [\boldsymbol{\Psi}_\mathrm{R}^\theta]_{q,\beta} [\mathbf{H}_\mathrm{a}^{\theta\phi}]_{\beta,\alpha} [\boldsymbol{\Psi}_\mathrm{S}^\phi]_{p,\alpha}^* \eta_{\mathrm{S},p} \\
		&+ \sum_{\beta=1}^{n_\mathrm{R}} \sum_{\alpha=1}^{n_\mathrm{S}} \eta_{\mathrm{R},q} [\boldsymbol{\Psi}_\mathrm{R}^\phi]_{q,\beta} [\mathbf{H}_\mathrm{a}^{\phi\theta}]_{\beta,\alpha} [\boldsymbol{\Psi}_\mathrm{S}^\theta]_{p,\alpha}^* \eta_{\mathrm{S},p} \\
		&+ \sum_{\beta=1}^{n_\mathrm{R}} \sum_{\alpha=1}^{n_\mathrm{S}} \eta_{\mathrm{R},q} [\boldsymbol{\Psi}_\mathrm{R}^\phi]_{q,\beta} [\mathbf{H}_\mathrm{a}^{\phi\phi}]_{\beta,\alpha} [\boldsymbol{\Psi}_\mathrm{S}^\phi]_{p,\alpha}^* \eta_{\mathrm{S},p},
	\end{split}
\end{equation}
where $\boldsymbol{\Psi}_\mathrm{R}^\theta \in \mathbb{C}^{N_\mathrm{R} \times n_\mathrm{R}}$ and $\boldsymbol{\Psi}_\mathrm{S}^\theta \in \mathbb{C}^{N_\mathrm{S} \times n_\mathrm{S}}$ denote the modified Fourier harmonics with the horizontal polarization, while $\boldsymbol{\Psi}_\mathrm{R}^\phi \in \mathbb{C}^{N_\mathrm{R} \times n_\mathrm{R}}$ and $\boldsymbol{\Psi}_\mathrm{S}^\phi \in \mathbb{C}^{N_\mathrm{S} \times n_\mathrm{S}}$ denote the modified Fourier harmonics with vertical polarization. $\mathbf{H}_\mathrm{a}^{\theta\theta} \in \mathbb{C}^{n_\mathrm{R} \times n_\mathrm{S}}$, $\mathbf{H}_\mathrm{a}^{\phi\phi} \in \mathbb{C}^{n_\mathrm{R} \times n_\mathrm{S}}$, $\mathbf{H}_\mathrm{a}^{\theta\phi} \in \mathbb{C}^{n_\mathrm{R} \times n_\mathrm{S}}$, and $\mathbf{H}_\mathrm{a}^{\phi\theta} \in \mathbb{C}^{n_\mathrm{R} \times n_\mathrm{S}}$ denote the co-polarization and cross-polarization wavenumber-domain channel matrices. The details of these parameters are explained in the following paragraphs.

\textbf{Polarization of Antennas:} Firstly, the polarization characteristics of the antennas are modeled. The polarization of a specific antenna can be described by its antenna pattern~\cite{2015-Balanis-Antenna}. Therefore, we further modify the Fourier harmonics to take the polarization of the antennas into consideration. The modified Fourier harmonics with polarization at the receiver can be expressed as
\begin{equation} 
	\begin{split}
		[\boldsymbol{\Psi}_\mathrm{R}^{\theta}]_{q,\beta} =
		& \frac{1}{\sqrt{N_\mathrm{R}}} e^{j\left( \frac{2\pi l^x_\beta}{L_\mathrm{R}^x} r^x_q + \frac{2\pi l^y_\beta}{L_\mathrm{R}^y} r^y_q + \gamma_\mathrm{R}(l^x_\beta,l^y_\beta) r_q^z\right)} \\
		& \times F_{\mathrm{R},q}^{\theta}\left( \hat{\theta}_\mathrm{R}(l^x_\beta, l^y_\beta) , \hat{\phi}_\mathrm{R}(l^x_\beta, l^y_\beta) \right),
	\end{split}
\end{equation}
and
\begin{equation} 
	\begin{split}
		[\boldsymbol{\Psi}_\mathrm{R}^{\phi}]_{q,\beta} =
		& \frac{1}{\sqrt{N_\mathrm{R}}} e^{j\left( \frac{2\pi l^x_\beta}{L_\mathrm{R}^x} r^x_q + \frac{2\pi l^y_\beta}{L_\mathrm{R}^y} r^y_q + \gamma_\mathrm{R}(l^x_\beta,l^y_\beta) r_q^z\right)} \\
		& \times F_{\mathrm{R},q}^{\phi}\left( \hat{\theta}_\mathrm{R}(l^x_\beta, l^y_\beta) , \hat{\phi}_\mathrm{R}(l^x_\beta, l^y_\beta) \right),
	\end{split}
\end{equation}
where $F_{\mathrm{R},q}^{\theta} \left( \theta_\mathrm{R}, \phi_\mathrm{R} \right)$ and $F_{\mathrm{R},q}^{\phi} \left( \theta_\mathrm{R}, \phi_\mathrm{R} \right)$ denote the embedded element directivity patterns in the horizontal and the vertical polarization. Similarly, the modified Fourier harmonics at the transmitter can be expressed as
\begin{equation} 
	\begin{split}
		[\boldsymbol{\Psi}_\mathrm{S}^{\theta}]_{p,\alpha} =
		& \frac{1}{\sqrt{N_\mathrm{S}}} e^{j\left( \frac{2\pi m^x_\alpha}{L_\mathrm{S}^x} s^x_p + \frac{2\pi m^y_\alpha}{L_\mathrm{S}^y} s^y_p + \gamma_\mathrm{S}(m^x_\alpha,m^y_\alpha) s_p^z\right)} \\
		& \times F_{\mathrm{S},p}^{\theta}\left( \hat{\theta}_\mathrm{S}(m^x_\alpha, m^y_\alpha) , \hat{\phi}_\mathrm{S}(m^x_\alpha, m^y_\alpha) \right),
	\end{split}
\end{equation}
and
\begin{equation} 
	\begin{split}
		[\boldsymbol{\Psi}_\mathrm{S}^{\phi}]_{p,\alpha} =
		& \frac{1}{\sqrt{N_\mathrm{S}}} e^{j\left( \frac{2\pi m^x_\alpha}{L_\mathrm{S}^x} s^x_p + \frac{2\pi m^y_\alpha}{L_\mathrm{S}^y} s^y_p + \gamma_\mathrm{S}(m^x_\alpha,m^y_\alpha) s_p^z\right)} \\
		& \times F_{\mathrm{S},p}^{\phi}\left( \hat{\theta}_\mathrm{S}(m^x_\alpha, m^y_\alpha) , \hat{\phi}_\mathrm{S}(m^x_\alpha, m^y_\alpha) \right),
	\end{split}
\end{equation}
where $F_{\mathrm{S},p}^{\theta} \left( \theta_\mathrm{S}, \phi_\mathrm{S} \right)$ and $F_{\mathrm{S},p}^{\phi} \left( \theta_\mathrm{S}, \phi_\mathrm{S} \right)$ denote the embedded element directivity patterns in the horizontal and the vertical polarization for the $p$-th antenna at the transmitter. 

\textbf{Polarization of Propagation Environment:} Secondly, the polarization characteristics of the propagation environment is modeled. Similar to~\cite{2020-Standard-38901}, we involve random phase shifts and cross polarization power ratios (XPR) to model the polarization characteristics of the propagation environment. For the co-polarization wavenumber-domain channels, a phase shift is added to account for the polarization distortion by the propagation environment, which can be expressed as
\begin{equation} 
	[\mathbf{H}_{\mathrm{a}}^{\theta\theta}]_{\beta, \alpha} = [\mathbf{H}_{\mathrm{a}}]_{\beta, \alpha} \times e^{j\Phi^{\theta\theta}_{\beta, \alpha}},
\end{equation}
and
\begin{equation} 
	[\mathbf{H}_{\mathrm{a}}^{\phi\phi}]_{\beta, \alpha} = [\mathbf{H}_{\mathrm{a}}]_{\beta, \alpha} \times e^{j\Phi^{\phi\phi}_{\beta, \alpha}}.
\end{equation}
Otherwise, the cross-polarization wavenumber-domain channels can be expressed as
\begin{equation} 
	[\mathbf{H}_{\mathrm{a}}^{\theta\phi}]_{\beta, \alpha} = [\mathbf{H}_{\mathrm{a}}]_{\beta, \alpha} \times e^{j\Phi^{\theta\phi}_{\beta, \alpha}} \times \sqrt{\kappa^{-1}_{\beta, \alpha}},
\end{equation}
and
\begin{equation} 
	[\mathbf{H}_{\mathrm{a}}^{\phi\theta}]_{\beta, \alpha} = [\mathbf{H}_{\mathrm{a}}]_{\beta, \alpha} \times e^{j\Phi^{\phi\theta}_{\beta, \alpha}} \times \sqrt{\kappa^{-1}_{\beta, \alpha}},
\end{equation}
where $\Phi^{\theta\theta}_{\beta, \alpha}$, $\Phi^{\phi\phi}_{\beta, \alpha}$, $\Phi^{\phi\theta}_{\beta, \alpha}$, and $\Phi^{\theta\phi}_{\beta, \alpha}$ are random phase shifts following the uniform distribution within $[-\pi, \pi]$. $\kappa_{\beta, \alpha}$ denotes the XPR of the propagation environment, which follows the log-normal distribution $\kappa_{\beta, \alpha} = 10^{X_{\beta, \alpha} / 10}$ with $X_{\beta, \alpha} \sim \mathcal{N}(\mu_{\mathrm{XPR}}, \sigma^2_\mathrm{XPR})$.

\textbf{Matrix Formulation:} Finally, we transform the item-wise channel model into a matrix form. The item-wise channel model in~\eqref{equ-chn-pol-item} can be further expressed as
\begin{equation} 
	\begin{split}
		\mathbf{H}^\mathrm{pol} = 
		& \boldsymbol{\Gamma}_\mathrm{R} \boldsymbol{\Psi}_\mathrm{R}^{\theta} {\mathbf{H}}_\mathrm{a}^{\theta\theta} {\boldsymbol{\Psi}_\mathrm{S}^\theta}^\mathrm{H} \boldsymbol{\Gamma}_\mathrm{S} + \boldsymbol{\Gamma}_\mathrm{R} \boldsymbol{\Psi}_\mathrm{R}^{\theta} {\mathbf{H}}_\mathrm{a}^{\theta\phi} {\boldsymbol{\Psi}_\mathrm{S}^\phi}^\mathrm{H} \boldsymbol{\Gamma}_\mathrm{S}\\
		& + \boldsymbol{\Gamma}_\mathrm{R} \boldsymbol{\Psi}_\mathrm{R}^{\phi} {\mathbf{H}}_\mathrm{a}^{\phi\theta} {\boldsymbol{\Psi}_\mathrm{S}^\theta}^\mathrm{H} \boldsymbol{\Gamma}_\mathrm{S} + \boldsymbol{\Gamma}_\mathrm{R} \boldsymbol{\Psi}_\mathrm{R}^{\phi} {\mathbf{H}}_\mathrm{a}^{\phi\phi} {\boldsymbol{\Psi}_\mathrm{S}^\phi}^\mathrm{H} \boldsymbol{\Gamma}_\mathrm{S} \\
		= & \boldsymbol{\Gamma}_\mathrm{R} 
		\begin{bmatrix} \boldsymbol{\Psi}_\mathrm{R}^{\theta}, \boldsymbol{\Psi}_\mathrm{R}^{\phi}  \end{bmatrix}
		\begin{bmatrix} \mathbf{H}_\mathrm{a}^{\theta\theta} & \mathbf{H}_\mathrm{a}^{\theta\phi} \\ \mathbf{H}_\mathrm{a}^{\phi\theta} & \mathbf{H}_\mathrm{a}^{\phi\phi} \end{bmatrix}
		\begin{bmatrix} \boldsymbol{\Psi}_\mathrm{S}^{\theta}, \boldsymbol{\Psi}_\mathrm{S}^{\phi} \end{bmatrix}^{\mathrm{H}} \boldsymbol{\Gamma}_\mathrm{S}.
	\end{split}
\end{equation}
Therefore, the final channel matrix can be derived as
\begin{equation} \label{equ-chn-pol-mat}
	\mathbf{H}^{\mathrm{pol}} = \boldsymbol{\Gamma}_\mathrm{R} \boldsymbol{\Psi}_\mathrm{R}^{\mathrm{pol}} {\mathbf{H}}_\mathrm{a}^{\mathrm{pol}} {\boldsymbol{\Psi}_\mathrm{S}^{\mathrm{pol}}}^\mathrm{H} \boldsymbol{\Gamma}_\mathrm{S},
\end{equation}
where $\boldsymbol{\Psi}_\mathrm{R}^{\mathrm{pol}} = [ \boldsymbol{\Psi}_\mathrm{R}^{\theta}, \boldsymbol{\Psi}_\mathrm{R}^{\phi} ]$, $\boldsymbol{\Psi}_\mathrm{S}^{\mathrm{pol}} = [ \boldsymbol{\Psi}_\mathrm{S}^{\theta}, \boldsymbol{\Psi}_\mathrm{S}^{\phi} ]$, and $\mathbf{H}_\mathrm{a}^{\mathrm{pol}} = \begin{bmatrix} \mathbf{H}_\mathrm{a}^{\theta\theta} & \mathbf{H}_\mathrm{a}^{\theta\phi} \\ \mathbf{H}_\mathrm{a}^{\phi\theta} & \mathbf{H}_\mathrm{a}^{\phi\phi} \end{bmatrix}$. 

As a result, not only the non-isotropic characteristics of the propagation environment, the antenna pattern distortion, the antenna efficiency, but also the polarization characteristics of the antennas and the propagation environment are all taken into consideration in the extended channel model.

\section{Measurement Setup} \label{Sec-Setu}
In order to evaluate the extended EM-compliant channel model and implement a realistic performance evaluation for holographic MIMO systems, an experiment is conducted to measure the real-world channel of holographic MIMO systems. To the best of our knowledge, it is the first attempt to measure the channel of a holographic MIMO system.

\begin{figure}[!t]\centering
	\includegraphics[width=0.47\textwidth]{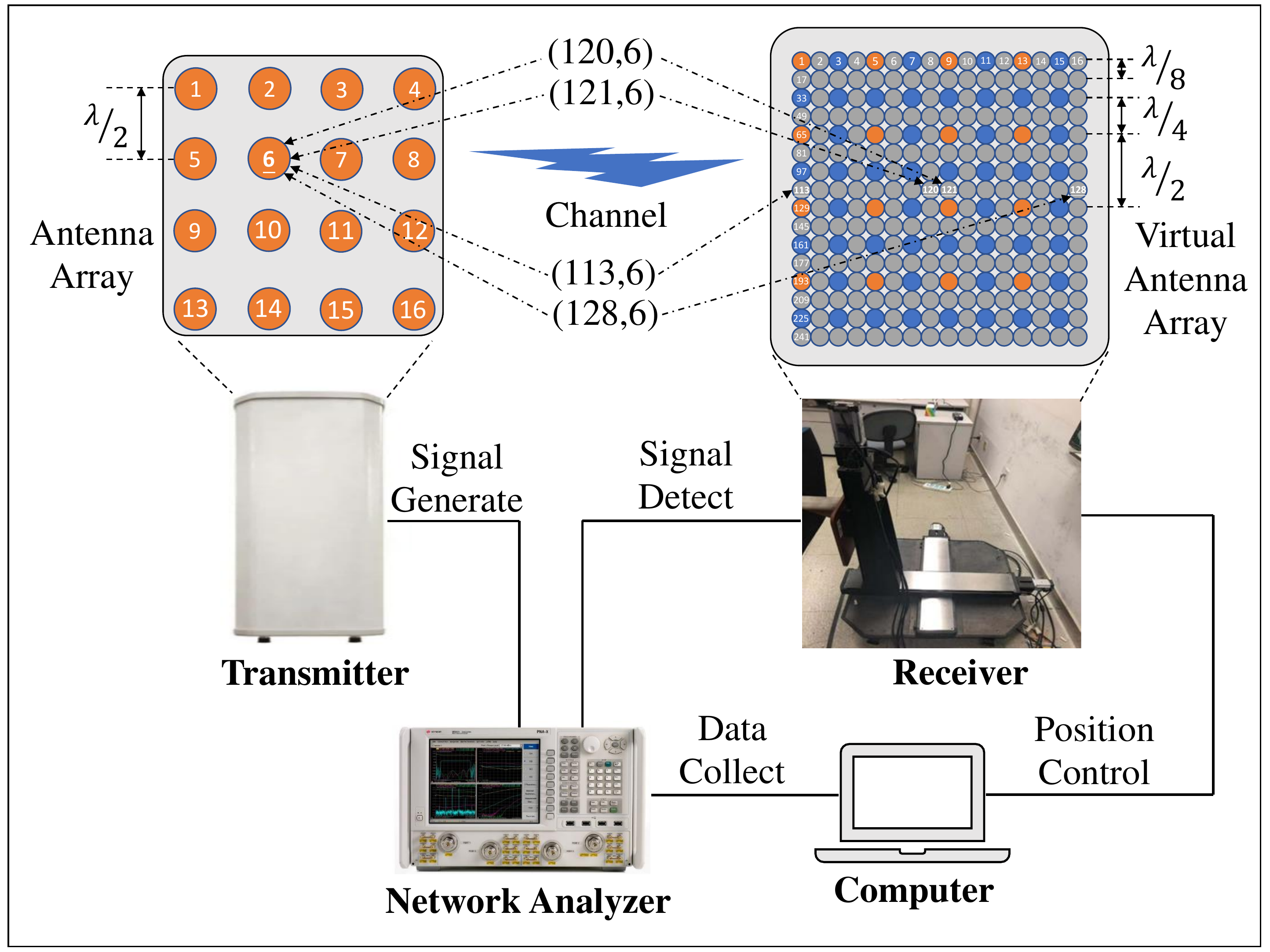}
	\caption{Schematic diagram of the measurement equipment.}
	\label{tx_rx}
\end{figure}

A schematic diagram of the measurement equipment is shown in Fig.~\ref{tx_rx}. In the experiment, the dense antenna array of holographic MIMO is realized by a virtual antenna array. A discone antenna is used to achieve an omnidirectional pattern and the position of it is controlled by an electrical machine. Through computer programming, the antenna can be moved to different positions to construct a virtual dense array with arbitrary element spacings. In the experiment, the virtual antenna array is equipped at the receiver to realize a holographic MIMO array with spacing $\Delta_\mathrm{R}^x = \Delta_\mathrm{R}^y \in \{\lambda /8,\lambda /4,\lambda /2\}$. At the transmitter, a conventional antenna array with $N_{\mathrm{S}} = 16$ antennas and spacings $\Delta_\mathrm{S}^x = \Delta_\mathrm{S}^y = \lambda /2$ is used. It is composed of patch antennas whose half power beam width is 70$^{\circ}$. A calibrated network analyzer is used to measure the channel and a computer is utilized to collect the measurement results. The center frequency is $f_c = 4.7$ GHz and the bandwidth is $200$~MHz from $4.6$~GHz to $4.8$~GHz with $1023$ samples. 

\begin{figure}[!t]\centering
	\includegraphics[width=0.47\textwidth]{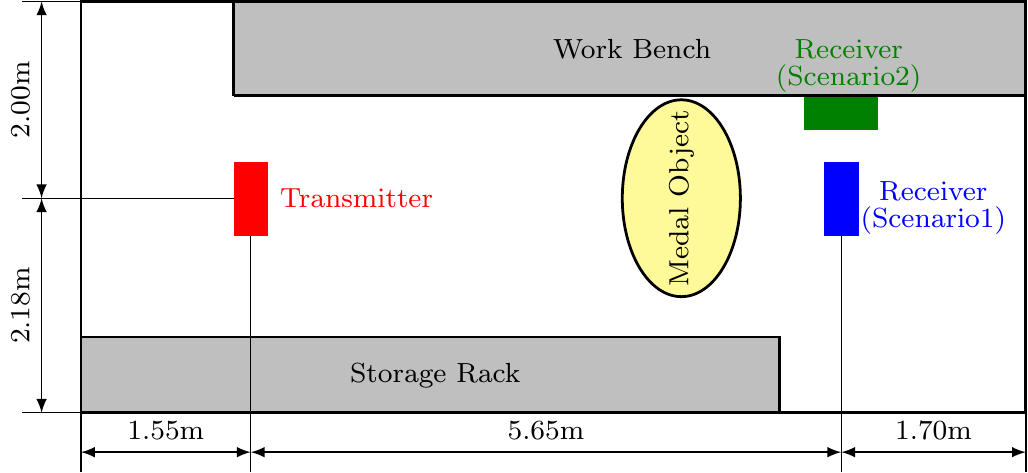}
	\caption{Schematic diagram of the measurement environment.}
	\label{ex_scenario}
\end{figure}

The experiment is performed in an indoor environment where the line-of-sight path is blocked by a metal object.
Many scatters are present to create a rich scattering environment.
The schematic diagram of the measurement environment is shown in Fig.~\ref{ex_scenario}.
We consider two scenarios.
In the first scenario, the virtual receive array plane is perpendicular to the transmitter, while in the second scenario, the virtual receive array plane is parallel to the transmit array plane.

\section{Measurement Results and Evaluations}  \label{Sec-Simu}

In this section, we use the measurement results to evaluate the performance of an indoor holographic MIMO system. The measurement results are shown in Section~\ref{Mea-Res} and corresponding performance evaluations are provided in Section~\ref{Mea-Eva}. Because the dense array of holographic MIMO is implemented virtually, the antenna efficiency loss is not accounted for in the measurement. Finally, the performance evaluations with antenna efficiency loss are provided in Section~\ref{Mea-Mul}.

\subsection{Channel Measurement Results}\label{Mea-Res}

\begin{figure}[!t]
	\centering
	\subfloat[ \label{S_a}]{%
		\includegraphics[width=0.9\linewidth]{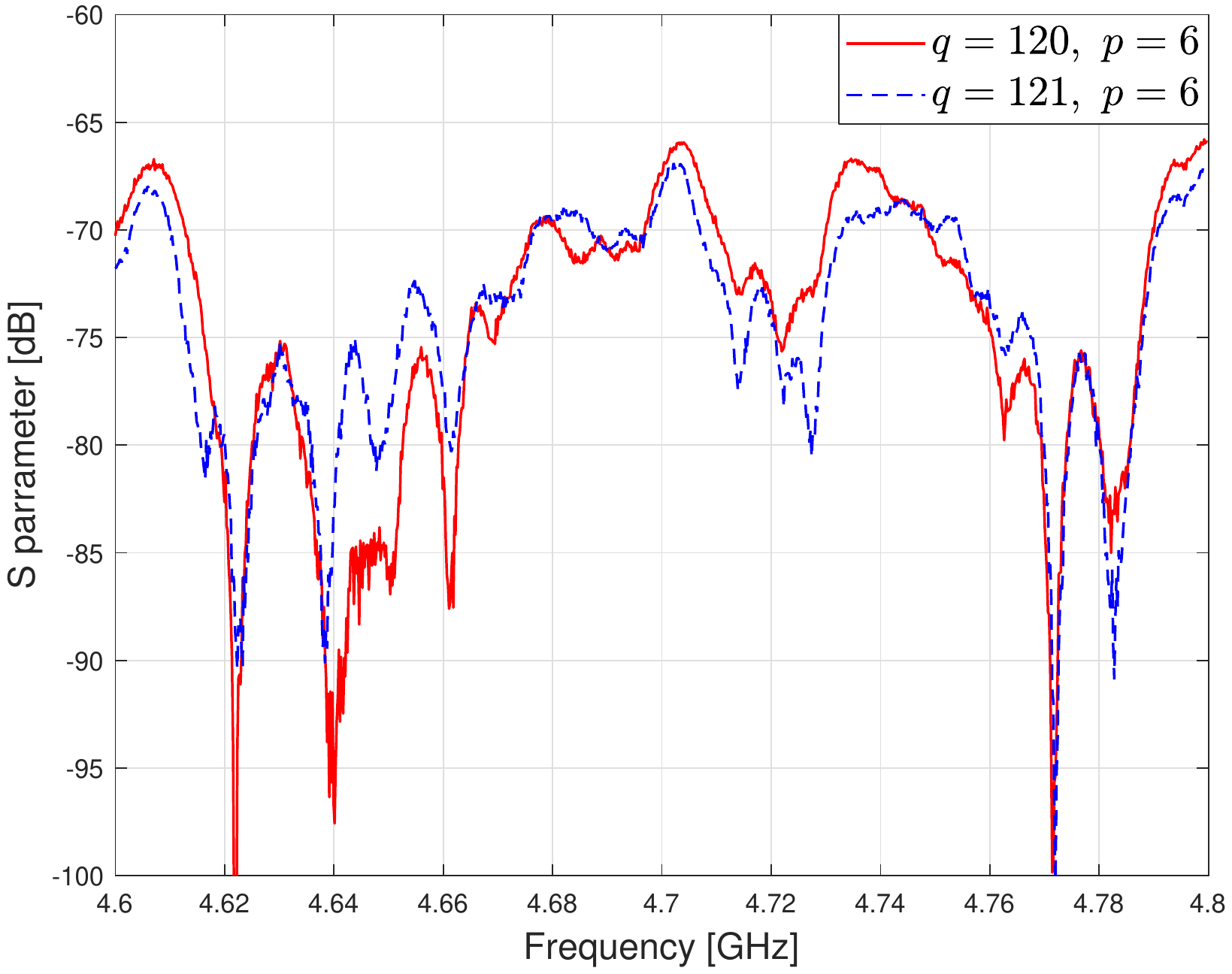}}
	\quad
	\subfloat[ \label{S_b}]{%
		\includegraphics[width=0.9\linewidth]{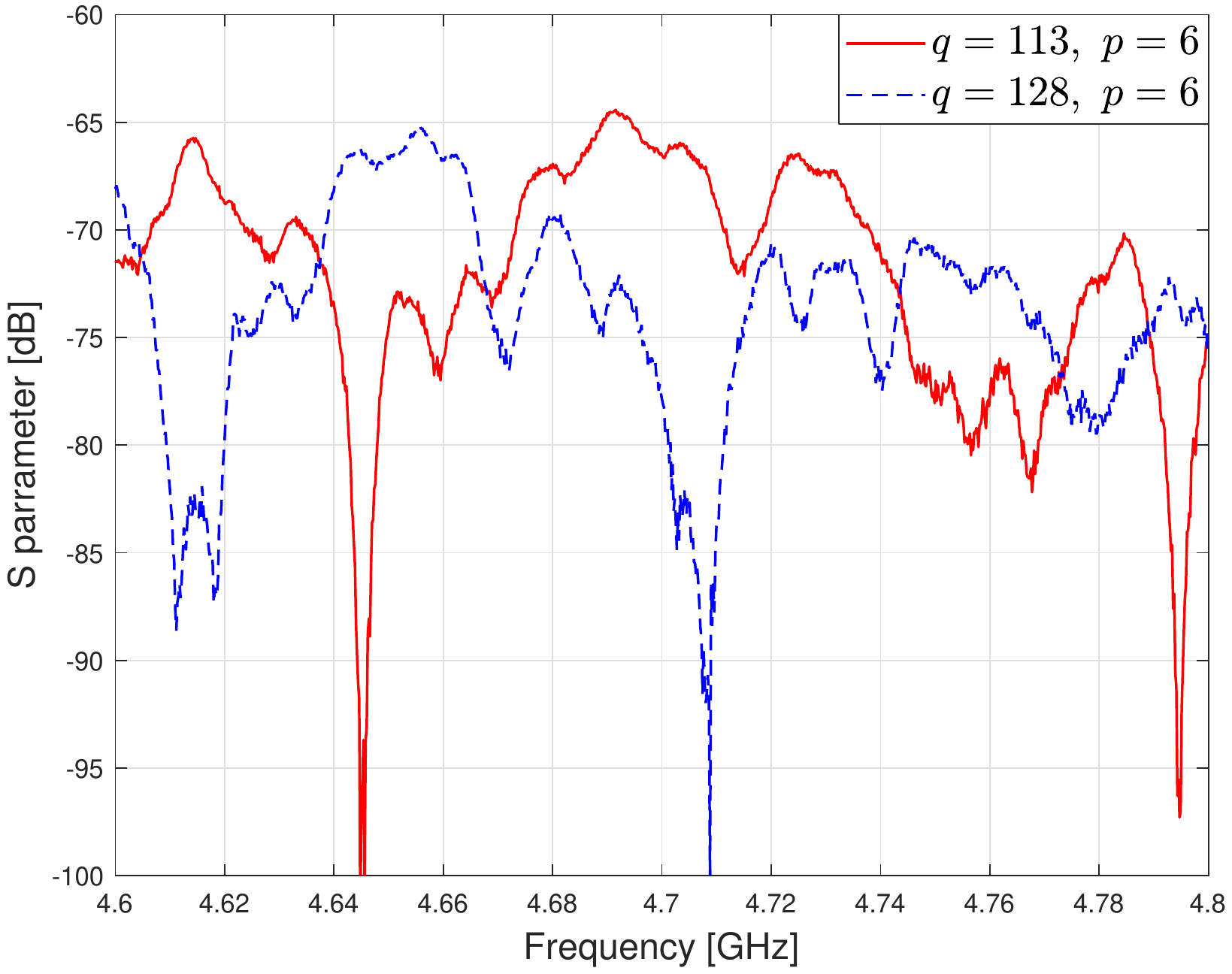}}
	\caption{Channel response between $q$-th antenna at the receiver and $p$-th antenna at the transmitter in Scenario 2. (a) $q=120,121$, $p=6$; (b) $q=113,128$, $p=6$.}
	\label{S_param} 
\end{figure}

After measurement, a channel matrix $\mathbf{H}^\mathrm{pol}$ with size $N_{\mathrm{R}} \times N_{\mathrm{S}}$ can be derived. Here we plot several channel measurement results to show the correlation of antennas at different position. We use the pair $(q,p)$ to represent the $q$-th antenna in the virtual receive array and the $p$-th antenna in the transmit array.

The channel responses corresponding to $(120,6)$ and $(121,6)$ transceiver pairs are shown in Fig.~\ref{S_a}. In these two pairs, the transmit antennas are the same and the receive antennas are adjacent, we can observe that the channel responses are quite similar.
Instead, if we choose transceiver pairs whose receiver elements are not adjacent, e.g., $(113,6)$ and $(128,6)$, the results are shown in Fig.~\ref{S_b}. Since their receiver elements are separated by $2 \lambda$, we can find that the red line differs from the blue line in the spectrum, showing a low correlation compared with the results in Fig.~\ref{S_a}.
The difference between these two figures shows the effect of spatial coherence. In an array, adjacent elements are more likely to sense the channel inside the same cluster, and thus the correlation of their channel responses are stronger.

\subsection{Performance Evaluation without Antenna Efficiency Loss}\label{Mea-Eva}

\begin{figure}[!t]
	\centering
	\subfloat[\label{C_a}]{%
		\includegraphics[width=0.95\linewidth]{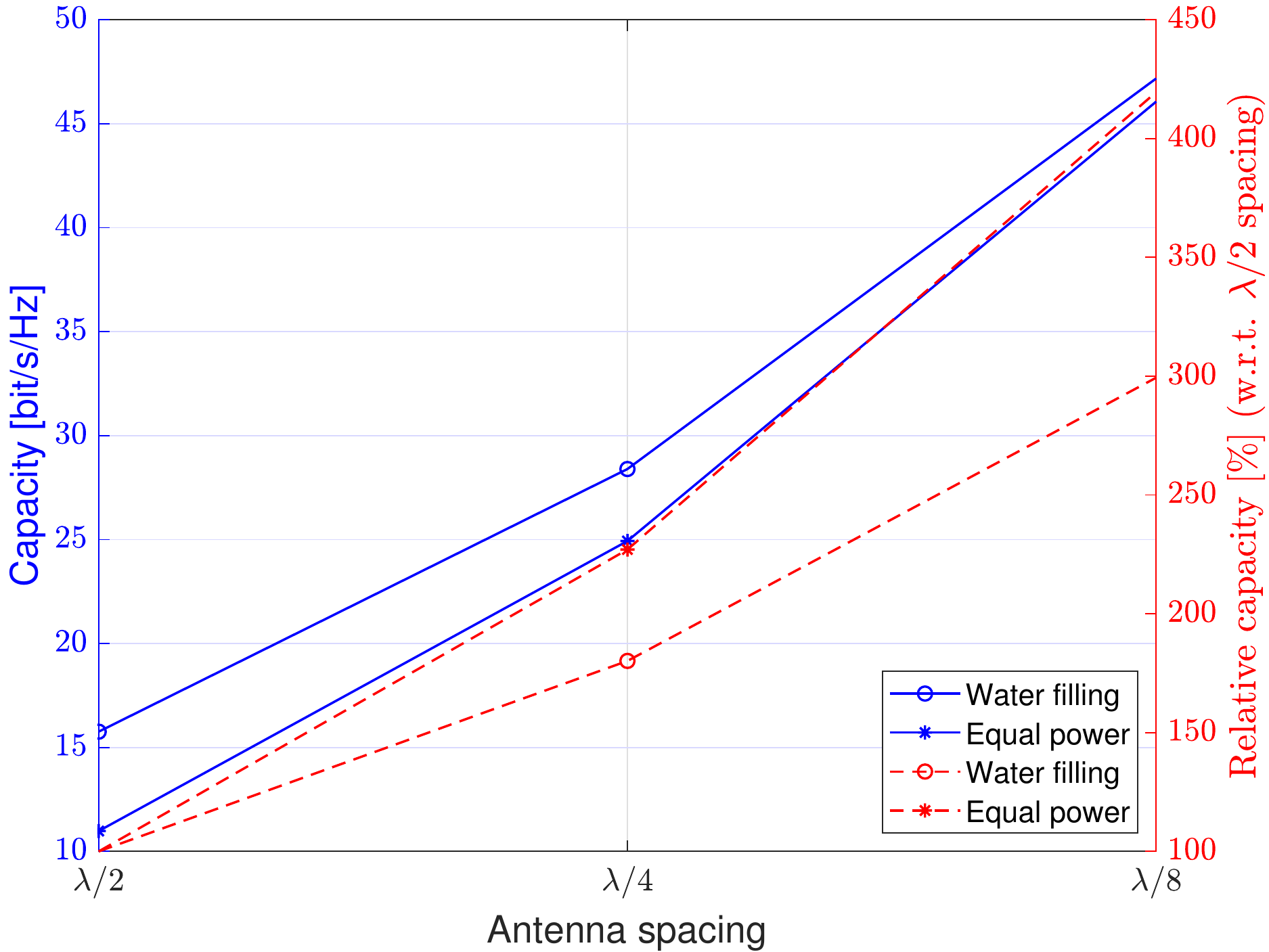}}
	\quad
	\subfloat[\label{C_b}]{%
		\includegraphics[width=0.95\linewidth]{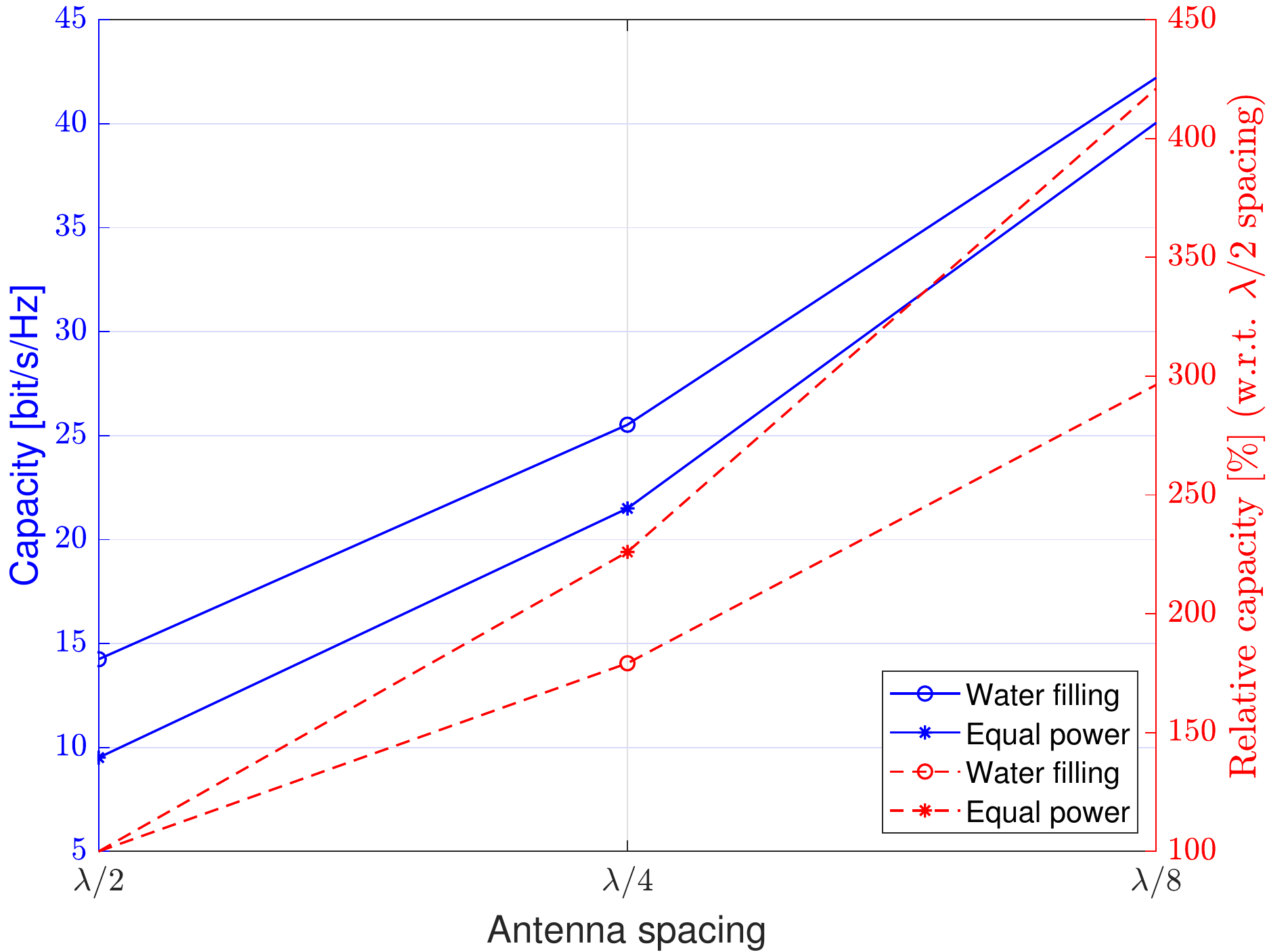}}
	\caption{Channel capacity and relative capacity with different antenna spacings. (a) Scenario 1; (b) Scenario 2.}
	\label{Cap_ideal} 
\end{figure}

Once the channel matrix $\mathbf{H}^\mathrm{pol}$ is obtained, we can evaluate the channel capacity based on the measurement results. The variation of channel capacity with different element spacings in both scenarios are shown in Fig~\ref{Cap_ideal}. In these figures, the antenna spacings are $\Delta_\mathrm{R}^x = \Delta_\mathrm{R}^y \in \{\lambda /8,\lambda /4,\lambda /2\}$ and the corresponding numbers of antennas are $N_{\mathrm{R}} \in \{256,64,16\}$ at the receiver. The signal to noise ratio (SNR) is set to 0 dB. Both the water filling and the equal power allocation strategies are adopted to evaluate the channel capacity performance. The blue lines correspond to the channel capacity, while the red lines correspond to the relative capacity with respect to the case with $\Delta_\mathrm{R}^x = \Delta_\mathrm{R}^y = \lambda /2$ spacing.

From the results in Fig.~\ref{Cap_ideal}, we can see that the spatial oversampling of holographic MIMO is able to increase the channel capacity. Using equal power allocation strategy, a four times spatial oversampling with $\Delta_\mathrm{R}^x = \Delta_\mathrm{R}^y = \lambda / 4$ can offer about $120\%$ capacity gain, and a $16$ times oversampling with $\Delta_\mathrm{R}^x = \Delta_\mathrm{R}^y = \lambda / 8$ provides more than $300\%$ capacity gain. While using the water filling strategy, the corresponding capacity gains are about $80\%$ and $200\%$.
Therefore, the capacity enhancement capability of holographic MIMO stated in the previous research works~\cite{2022-Pizzo-HoloChannel, 2022-Tengjiao-Holo, 2021-Linglong-CAP} is verified by practical experiment. It is worth noting that the antenna efficiency loss at the receiver is not taken into consideration in the measurement because the dense array is realized virtually, which means $\eta_{\mathrm{R}, q} = 1$. In the next subsection, the antenna efficiency loss is further considered in analyzing the capacity of a holographic MIMO system.

\begin{figure*}[!t] 
	\centering
	\subfloat[ ]{%
		\includegraphics[width=0.47\linewidth]{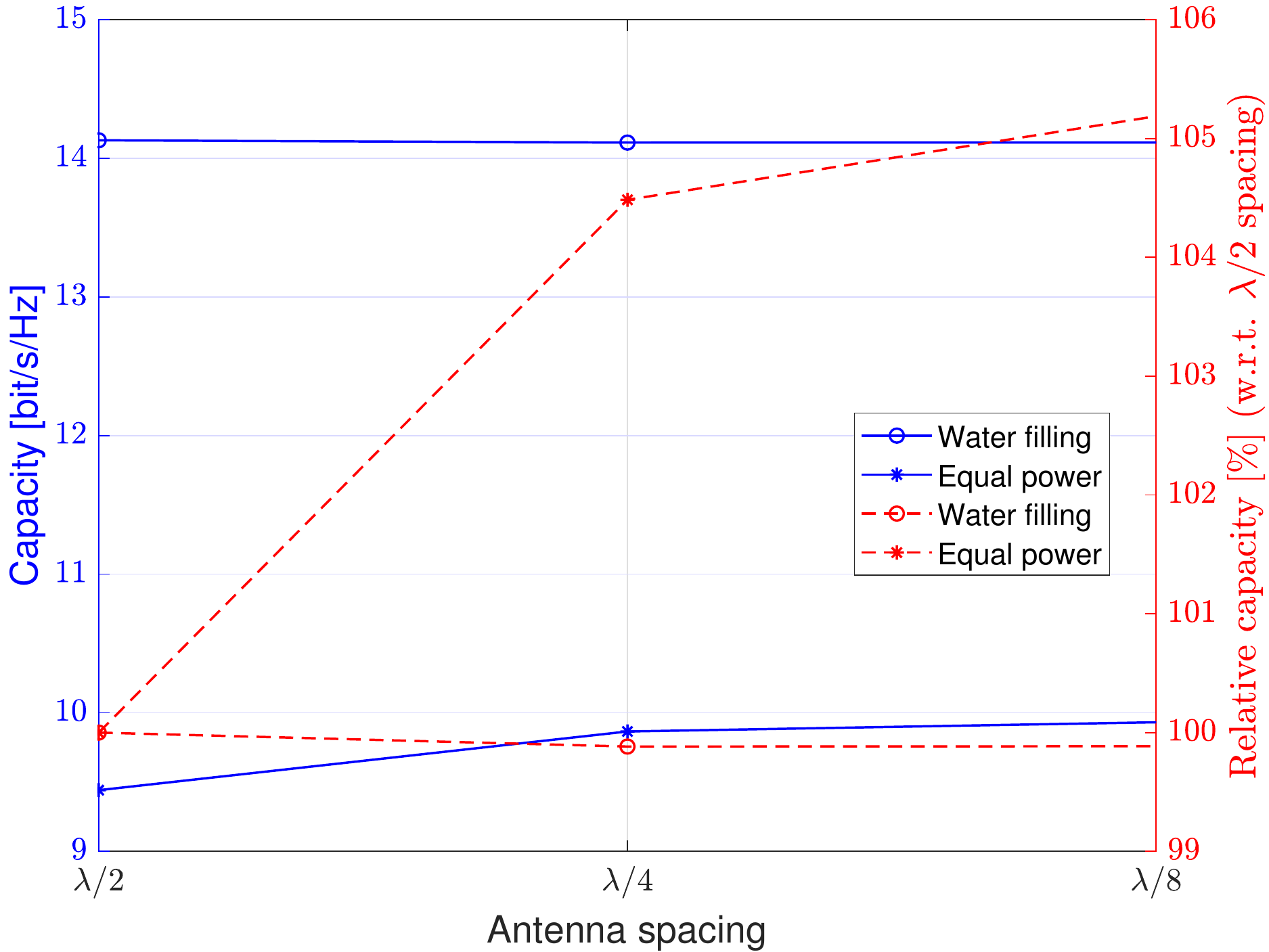}}
	\quad
	\subfloat[ ]{%
		\includegraphics[width=0.47\linewidth]{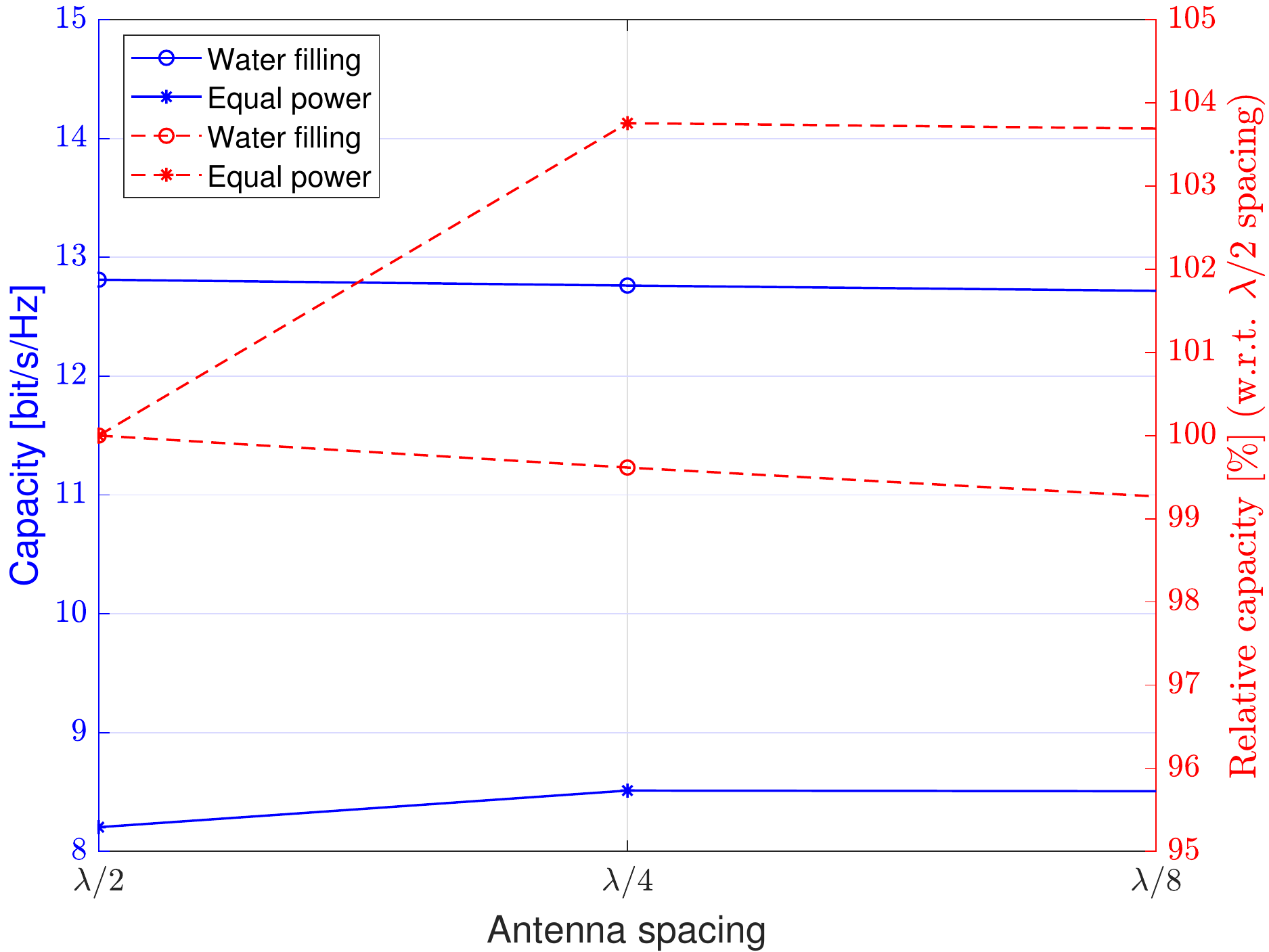}}
	\caption{Channel capacity and relative capacity with antenna efficiency loss. (a) Scenario 1; (b) Scenario 2.}
	\label{Cap_real} 
\end{figure*}

\subsection{Performance Evaluation with Antenna Efficiency Loss}\label{Mea-Mul}
In a practical dense antenna array with small element spacing, the efficiency of the antenna elements will decrease because of the mutual coupling among them. In~\cite{1964-Hannan-Limit}, a relationship between the antenna efficiency and the element spacing is established for a dense array, which is called Hannan's element efficiency. According to~\cite{1964-Hannan-Limit}, for a practical dense array at the receiver, the efficiency of the antenna element can be estimated as
\begin{equation} \label{equ-Hannan}
    \eta_{\mathrm{R},q} \approx \frac{\pi \Delta_\mathrm{R}^x \Delta_\mathrm{R}^y}{\lambda^2},
\end{equation}
which means that the element efficiency is proportional to the area allocated to the element. It implies that when the spacing of antenna element is small ($\Delta_\mathrm{R}^x \Delta_\mathrm{R}^y < \lambda ^2 / \pi$), the element efficiency cannot reach $1$, and it will decrease as the antenna elements are placed closer.

Using the efficiency estimation in~\eqref{equ-Hannan}, we modify the channel measurement results and evaluate the channel capacity of holographic MIMO systems. The results are shown in Fig.~\ref{Cap_real}. It can be seen that in both scenarios, the channel capacities will not keep increasing with more antenna elements and smaller element spacings. Using the equal power allocation strategy, a $16$ times oversampling with $\Delta_\mathrm{R}^x = \Delta_\mathrm{R}^y = \lambda / 8$ can only provide $4\%$ capacity gain. While using the water filling strategy, the channel capacities even slightly decrease. The reason behind this is that the array gain and multiplexing gain by deploying more antenna elements are reduced by the decrease of the antenna efficiency.

From the above analyses, we can find that although the channel correlation increases with smaller antenna spacings, the spatial oversampling of holographic MIMO is able to offer an obvious capacity enhancement. However, the antenna efficiency loss due to mutual coupling will greatly decrease the capacity gain, which is one of the most important challenges for a practical holographic MIMO system. Therefore, designing a dense antenna array with element efficiency above the Hannan's efficiency scaling law will be the promising ways to exploit the benefit of spatial oversampling for the holographic MIMO systems.

\section{Conclusion} \label{Sec-Conc}
In this paper, an extended EM-compliant channel model is proposed for holographic MIMO systems, which takes the non-isotropic characteristics of the propagation environment, the antenna pattern distortion, the antenna efficiency, and the polarization into account. An experiment is also conducted to measure the channel of an indoor holographic MIMO system through virtual antenna arrays. It is demonstrated through experiments for the first time that the spatial oversampling of holographic MIMO is able to increase the capacity of a wireless communication system significantly. However, the antenna efficiency is the most crucial challenge preventing us from getting the capacity improvement. 

\bibliographystyle{IEEEtran}
\bibliography{MyBib}

\vfill

\end{document}